\documentclass[aps,pre,superscriptaddress,twocolumn,longbibliography,bibtex]{revtex4-1}

\usepackage[utf8]{inputenc}  
\usepackage[T1]{fontenc}  

\usepackage[tbtags]{amsmath}
\usepackage{amssymb}
\usepackage{mathtools}

\usepackage{float}

\usepackage{xcolor}

\definecolor{myOrange}{rgb}{1,0.5,0}
\definecolor{myRed}{rgb}{0.8, 0.2, 0}
\definecolor{mygreen}{rgb}{0, 0.7, 0}

\usepackage[colorlinks=true,urlcolor=blue,linkcolor=blue,citecolor=blue]{hyperref}  

\renewcommand{\Re}{\mathrm{Re}}

\newcommand{\dv}[1]{\frac{\mathrm{d}}{\mathrm{d}#1}}

\newcommand{\dd}{\mathrm{d}}

\newcommand{\abs}[1]{\lvert #1 \rvert}

\newcommand{\expsup}[1]{\mathrm{e}^{#1}}

\newcommand{\genmatDrate}{\genmatDrate_{\Doobletter}}

\newcommand{\Eorder}{m}
\newcommand{\dens}{\rho}
\newcommand{\densxt}{\dens(\pos,\tv)}
\newcommand{\meandens}{\rho_0}

\newcommand{\denspxt}{\delta\rho(\pos,\tv)}
\newcommand{\tv}{t}
\newcommand{\pos}{x}
\newcommand{\diffcoef}{D}
\newcommand{\mobility}{\sigma}

\newcommand{\Ed}{\epsilon} 
\newcommand{\Epampl}{\lambda} 
\newcommand{\Epamplcrit}[1]{\lambda_{\mathrm{c}}^{(#1)}} 
\newcommand{\twarg}{u}
\newcommand{\twargt}{\tilde{\twarg}}

\newcommand{\fcoef}[1]{C_{#1}} 
\newcommand{\fcoeft}[1]{\fcoef{#1}(\tv)} 
\newcommand{\opar}{z} 
\newcommand{\oparm}{\opar_{\Eorder}} 

\newcommand{\denstw}[1]{f_{#1}}

\newcommand{\denstwt}{\tilde{\denstw{}}}
\newcommand{\denstwtprime}{\tilde{\denstw{}'}}

\newcommand{\meancurrent}{J}
\newcommand{\meancurrentflat}{\meancurrent_0}
\newcommand{\opararg}{\varphi}
\newcommand{\oparargm}{\varphi_{\Eorder}}
\newcommand{\oparargmini}{\opararg_{\Eorder, 0}}

\newcommand{\titulo}{Programmable time crystals from higher-order packing fields}

\begin{document}

\title{\titulo}
\author{R. Hurtado-Guti\'errez}
\affiliation{Departamento de Electromagnetismo y F\'isica de la Materia, Universidad de Granada, 18071 Granada, Spain}
\affiliation{Institute Carlos I for Theoretical and Computational Physics, Universidad de Granada, 18071 Granada, Spain}
\author{C. P\'erez-Espigares}
\affiliation{Departamento de Electromagnetismo y F\'isica de la Materia, Universidad de Granada, 18071 Granada, Spain}
\affiliation{Institute Carlos I for Theoretical and Computational Physics, Universidad de Granada, 18071 Granada, Spain}
\author{P.I. Hurtado}
\affiliation{Departamento de Electromagnetismo y F\'isica de la Materia, Universidad de Granada, 18071 Granada, Spain}
\affiliation{Institute Carlos I for Theoretical and Computational Physics, Universidad de Granada, 18071 Granada, Spain}
\date{\today}

\begin{abstract}
Time crystals are many-body systems that spontaneously break time-translation symmetry, and thus exhibit long-range spatiotemporal order and robust periodic motion. Recent results have shown that coupling an external packing field to density fluctuations in driven diffusive fluids can trigger a transition to a time-crystal phase. Here we exploit this mechanism to engineer and control on demand programmable continuous time crystals characterized by an arbitrary number of rotating condensates, which can be further enhanced with higher-order modes. We elucidate the underlying critical point, as well as general properties of the condensates density profiles and velocities, demonstrating a scaling property of higher-order traveling condensates in terms of first-order ones. We illustrate our findings by solving the hydrodynamic equations for various paradigmatic driven diffusive systems, obtaining along the way a number of remarkable results, as e.g. the possibility of explosive time crystal phases characterized by an abrupt, first-order-type transition. Overall, these results demonstrate the versatility and broad possibilities of this promising route to time crystals.
\end{abstract}

\maketitle

\section{Introduction}
\label{sec:1}
The concept of time crystal, first introduced by Wilczek and Shapere \cite{wilczek12a, shapere12a}, describes many-body systems that spontaneously break time-translation symmetry, a phenomenon that leads to persistent oscillatory behavior and fundamental periodicity in time  \cite{zakrzewski12a,richerme17a,yao18a, sacha18a,sacha20a}. The fact that a symmetry might appear broken comes as no surprise in general. Indeed spontaneous symmetry-breaking phenomena, where a system ground state can display fewer symmetries than the associated action, are common in nature. However, time-translation symmetry had resisted this picture for a long time, as it seemed fundamentally unbreakable. Progress made over the last decade has challenged this scenario showing that both continuous and discrete time-translation symmetries can be spontaneously broken, giving rise to the so-called continuous and discrete time crystals, respectively. In quantum settings, the former are prohibited in equilibrium short-ranged systems by virtue of a series of no-go theorems \cite{bruno13a,nozieres13a,watanabe15a,kozin19a}, which are however circumvented in nonequilibrium dissipative contexts allowing for continuous time crystals \cite{iemini18a,buca19aa,kessler19a,carollo20a1,carollo22a,Kongkhambut22a}. On the other hand, quantum discrete time crystals can emerge as a subharmonic response to a periodic (Floquet) driving \cite{else16a, moessner17a, yao17a, gong18a, gambetta19a, khemani2019brief, Lazarides20a, Else20a, zhang17a, choi17a, Rovny18a, smits18a, Autti18a, Sullivan20a, Kyprianidis21a, Randall21a, Mi22a, kessler21a, Kongkhambut21a}, while some classical systems have been also shown to exhibit time-crystalline order \cite{gambetta19b, yao20a, Heugel19a, Liu23a}. However, a general approach to engineer custom time-crystal phases remains elusive so far.

\begin{figure}[b!]
\includegraphics[width=0.95\linewidth]{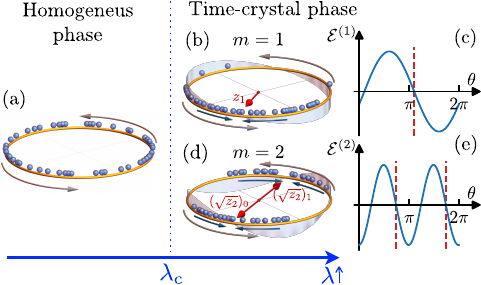}
\vspace{-1mm}
\caption{
(a) Sketch of a $1d$ driven particle system sustaining a net current with an homogeneous density structure on average. By switching on a $m$th-order packing field ${\cal E}^{(m)}(\theta)$ with strength $\lambda$ beyond a critical value [see (c),(e) and shaded curves in (b),(d)], an instability is triggered to a time-crystal phase characterized by the emergence of $\Eorder$ rotating particle condensates. The magnitude $\abs{\oparm}$ of the complex packing order parameter indicates the packing of particles around $m$ emergent localization centers, placed at the argument of $(\hspace{-3pt}\sqrt[m]{\oparm})_j$, with $j\in[0,m-1]$, and represented by the red arrows inside the ring in (b),(d). The convergent blue arrows around the ring signal the local direction of the packing field around the $m$ localization centers.
}
\label{fig1new}
\end{figure}

In this work we propose a general mechanism to build programmable continuous time crystals in driven diffusive fluids, grounded on hydrodynamics. This universal description of mesoscopic dynamics captures the physics of a broad family of systems, making our results widely applicable while adding a valuable tool to the repertoire of pattern formation control strategies. An experimental example where our theory would apply consists in an assembly of dispersed colloidal particles trapped in a rotating quasi-$1d$ ring periodic optical potential \cite{cereceda-lopez23a,cereceda-lopez23b,cereceda-lopez24a} or channel \cite{lutz04a,villada-balbuena21a}. This periodic channel can be seen as a $1d$ ring embedded in a $2d$ plane, see Fig. \ref{fig1new}. Our approach leverages the concept of \emph{packing field} \cite{hurtado-gutierrez20a,hurtado-gutierrez23a}, a mechanism inspired by the rare event statistics of some driven diffusive systems \cite{bertini05a, bodineau05a, bertini06a, derrida07a, hurtado11a, perez-espigares13a, hurtado14a, lazarescu15a,tizon-escamilla17b, shpielberg18a,perez-espigares19a}. This packing field $\mathcal{E}^{(1)}(\theta)$ acts by pushing particles that lag behind the center of mass of any emergent particle condensate, while simultaneously restraining those moving ahead, as sketched in Fig.~\ref{fig1new}.(b)-(c). This amplifies naturally-occurring fluctuations of the particles' spatial packing, a nonlinear feedback mechanism that eventually leads to a time crystal. From a mathematical perspective, the action of the packing field can be seen as a controlled excitation of the first Fourier mode of the density field around the instantaneous center of mass position \cite{hurtado-gutierrez20a,hurtado-gutierrez23a}, see Fig.~\ref{fig1new}.(b)-(c). This viewpoint immediately raises the natural question: What happens if we excite higher-order modes? Here we show how this leads to tailored and fully controllable continuous time-crystal phases in driven diffusive fluids, characterized by an arbitrary number $m$ of rotating condensates, as displayed in Fig \ref{fig1new}.(d)-(e), which can be further enhanced with higher-order modes. A local stability analysis of the governing hydrodynamic equations reveals the details of the transition to these intriguing time-crystal phases, along with general properties of the condensates density profiles and velocities. Using this hydrodynamic picture, we also demonstrate a scaling property of higher-order traveling condensates in terms of first-order ones. We illustrate these findings in several paradigmatic models, including the random walk fluid \cite{spohn12a}, the Kipnis-Marchioro-Presutti heat transport model \cite{kipnis82a}, the weakly asymmetric simple exclusion process for interacting particle diffusion \cite{spitzer70a,derrida98a}, and the Katz-Lebowitz-Spohn lattice gas \cite{katz84,hager01a,Krapivsky13a, baek17a}. Programmable time-crystal phases (i.e. with controlled number, shape, and velocity of the emerging condensates) are demonstrated and characterized in all these models, finding along the way a novel explosive time-crystal phase transition, controlled by the nonlinearity of transport coefficients. Altogether, these results show the versatility and broad possibilities of this promising route to custom time crystals.

\section{The packing-field route to time crystals}
\label{sec:2}
Our starting point is the hydrodynamic evolution equation for the density field $\rho(x,t)$ in a $1d$ periodic diffusive system driven by an external field $E_x[\rho]$ \cite{spohn12a},
\begin{equation}
\partial_t \rho = - \partial_x \Big[-D(\rho) \partial_x \rho + \sigma(\rho) E_x[\rho] \Big] \, ,
\label{eq:hydro_field}
\end{equation}
with $\pos \in [0, 1]$, and $\diffcoef(\dens)$ and $\mobility(\dens)$ the diffusivity and the mobility transport coefficients, respectively. The colloidal fluid example mentioned above is a particular instance of system governed by this type of transport equation, with $\theta=2\pi x$ the angular position in the $1d$ ring embedded in a $2d$ plane. The external field takes the form $E_x[\rho] = \epsilon +  \lambda \mathcal{E}_x^{(m)}[\rho]$, where $\epsilon$ is a constant driving that leads to a net current and $\lambda$ is the coupling constant to a $m$-th order packing field $\mathcal{E}_x^{(m)}[\rho]$ \cite{hurtado-gutierrez20a}. As discussed above, this packing field excites the $m$-th Fourier mode of the density field, i.e.
\begin{equation}
\mathcal{E}^{(m)}_x[\rho] = \frac{1}{\rho_0} \int_0^1 \dd y \, \rho(y,t) \sin\left(2\pi m (y-x) \right) \, ,
\label{eq:extfield_pairs_field}
\end{equation}
where $\rho_0=\int_0^1 \rho(x, t) \dd x$ is the conserved average density. To gain some physical insight on the action of $\mathcal{E}_x^{(m)}[\rho]$, we define now the complex $\Eorder$th-order packing order parameter \cite{hurtado-gutierrez20a,hurtado-gutierrez23a,daido1992},
\begin{equation}
z_m[\rho] = \frac{1}{\rho_0} \int_0^1 dx \, \rho(x,t) \, \textrm{e}^{\textrm{i}2\pi m x} \equiv \abs{z_m} \textrm{e}^{\textrm{i} \varphi_m}\, ,    
\label{eq:z_field}
\end{equation}
which is formally equivalent to the Kuramoto-Daido order parameter in the context of synchronization transitions \cite{kuramoto87a, pikovsky03a, acebron05a}, see \S\ref{sec:5} below. Its magnitude $\abs{\oparm}$ measures the packing of the density field around $\Eorder$ equidistant \emph{emergent localization centers} placed at angular positions $\phi_m^{(j)}=\arg[(\hspace{-3pt}\sqrt[m]{\oparm})_j]=(\varphi_m + 2\pi j)/m$, with $j\in[0,m-1]$. Using $z_m[\rho]$, the packing field \eqref{eq:extfield_pairs_field} can be simply rewritten as $\mathcal{E}_x^{(m)}[\rho] =  \abs{z_m} \sin(\varphi_m - 2\pi\pos \Eorder)$, see Fig. \ref{fig1new}.(c)-(e) with $\theta = 2 \pi x$. In this way, $\mathcal{E}^{(m)}_x[\rho]$ drives particles locally towards the $\Eorder$ emergent localization centers placed at $\phi_m^{(j)}\in[0,2\pi)$,  pushing particles that lag behind the closest localization center while restraining those moving ahead, with an amplitude proportional to the amount of local packing as measured by $\abs{\oparm}$, see Figs.~\ref{fig1new}.(b)-(e). This results in a nonlinear feedback mechanism that amplifies the local packing fluctuations naturally present in the system, resulting eventually in the emergence of $\Eorder$ traveling-wave condensates for large enough values of $\Epampl$, and exhibiting the fingerprints of spontaneous time-translation symmetry breaking.

\section{Hydrodynamic instability and condensate equivalence}
\label{sec:3}
To determine the critical threshold $\Epampl_c^{(\Eorder)}$ for this instability to happen, we first note that $\forall \Epampl$ the homogeneous density profile $\densxt = \meandens$ is a solution of Eq.~\eqref{eq:hydro_field}. We hence perform a linear stability analysis of this solution and introduce a small perturbation over the flat profile, $\densxt = \meandens + \denspxt$, with $\int_0^{1} d\pos~\denspxt = 0$ to conserve the global density. Linearizing Eq.~\eqref{eq:hydro_field} and expanding $\denspxt$ in Fourier modes, it can be shown (see Appendix  \ref{app1}) that the different modes decouple and their stability depend on a competition between the diffusion term and  the packing field, controlled by the coupling.  This results in the $m$-th Fourier mode becoming unstable when $\lambda > \lambda_c^{(\Eorder)}$, with
\begin{equation}
\lambda_c^{(\Eorder)} = 4\pi\Eorder \frac{\diffcoef(\meandens)\meandens}{\mobility(\meandens)} \, .
\label{lambc}
\end{equation}
The form of the resulting perturbation beyond the instability (see Appendix \ref{app1}) is compatible with the emergence of $m$ traveling-wave condensates, $\densxt =\rho_m(\omega_m t - 2\pi mx)$, moving periodically with an angular velocity $\omega_m= 2\pi m \sigma'(\rho_0) \epsilon$ right after the instability ($\lambda\gtrsim \lambda_c^{(\Eorder)}$). This instability breaks spontaneously the time-translation symmetry of the flat solution, thus giving rise to a continuous time crystal \cite{wilczek12a, shapere12a,zakrzewski12a,richerme17a,yao18a, sacha18a,sacha20a,hurtado-gutierrez20a}. Interestingly, the value of $\lambda_c^{(\Eorder)}$ increases with $m$, see Eq.~\eqref{lambc}, a reflection of the competition between diffusion and the packing field. Indeed, while the effect of diffusion, which tends to destroy the $m$ emergent condensates, scales as $m^2$ at the instability, the action of the packing field promoting the condensates scales as $m$, and therefore a stronger $\lambda$ is needed as $m$ increases to destabilize the flat solution. On the other hand, the excess of the averaged current $J=\tau^{-1}\int_0^\tau dt \int_0^1 dx~j(x,t)$ with respect to the homogeneous-phase average current $J_0=\sigma(\rho_0) \epsilon$ can be shown to be $J-J_0\propto \sigma''(\rho_0) \epsilon$ right after the instability, so the current will be larger or smaller than the homogeneous-phase current depending on the mobility curvature for density $\rho_0$. This highlights the relevance of transport coefficients in the system's reaction to the packing field, which enhances or lowers the current and the wave velocity depending on the mobility derivatives. 

\begin{figure}
\vspace{-5mm}
\includegraphics[width=\linewidth]{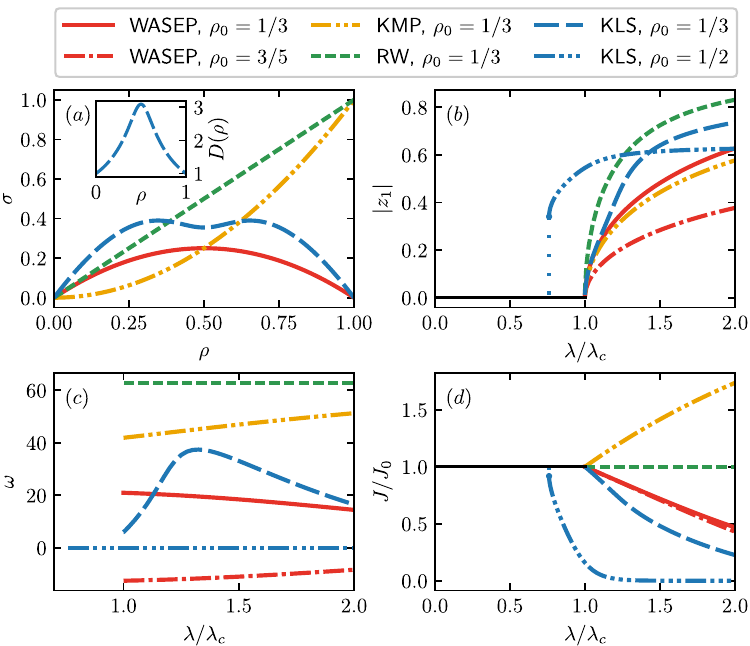}
\vspace{-5mm}
\caption{ 
(a) Mobility $\sigma(\rho)$ for the different models studied. Inset: diffusivity $D(\rho)$ in the KLS model. For $m=1$, $\epsilon=10$ and different $\rho_0$, the other panels display (b) the magnitude of the packing order parameter $|z_1|$, (c) the condensate velocity $\omega_1$, and (d) the average relative current $J/J_0$, as a function of $\Epampl/\Epampl_\mathrm{c}$ for the different models.
}
\label{fig2}
\end{figure}

A remarkable property of the emergent time-crystal phase is that the $m$-th-order traveling-wave solution can be built by \emph{gluing} together $m$ copies of the $m = 1$ solution after an appropriate rescaling of driving parameters. In particular, it can be shown (see Appendix \ref{app2}) that $\rho_m(\omega_m t - 2\pi mx) = \rho_1(m\omega_1 t - 2\pi mx)$, where $\rho_1(\omega_1 t - 2\pi x)$ is the traveling-wave solution of Eq.~\eqref{eq:hydro_field} of velocity $\omega_1$ for $m=1$ and parameters $\epsilon_1$ and $\lambda_1$, while $\rho_m(\omega_m t - 2\pi mx)$ is the corresponding traveling-wave solution of Eq.~\eqref{eq:hydro_field} of velocity $\omega_m=m\omega_1$ for arbitrary $m>1$ and parameters $\epsilon_m=m\epsilon_1$ and $\lambda_m=m\lambda_1$. This scaling property, valid for arbitrary transport coefficients, allows to collapse traveling-wave solutions for different orders $m$ and related driving parameters, reducing the range of possible solutions. Interestingly, a similar dynamical equivalence between the first-order and higher-order couplings has been reported for the particular case of the Kuramoto model \cite{delabays2019}, see \S\ref{sec:5} below for more details on this mathematical connection with synchronization phenomena.

\section{Examples}
\label{sec:4}
To illustrate our findings, we investigate now four paradigmatic driven diffusive models \cite{spohn12a} under the action of a packing field~\eqref{eq:extfield_pairs_field}, and which admit a hydrodynamic description of the form of Eq.~\eqref{eq:hydro_field}. These models are (i) the random walk (RW) fluid, which captures the diffusive motion of independent particles and is described by a diffusivity $D(\rho)=1/2$ and a linear mobility $\sigma(\rho)=\rho$ \cite{spohn12a}, (ii) the weakly asymmetric exclusion process (WASEP) that models particle diffusion under exclusion interactions, characterized by $D(\rho)=1/2$ and $\sigma(\rho)=\rho (1-\rho)$ \cite{spitzer70a,derrida98a}, (iii) the Kipnis-Marchioro-Presutti (KMP) model of heat transport \cite{kipnis82a}, with $D(\rho)=1/2$ and $\sigma(\rho)=\rho^2$, and (iv) the Katz-Lebowitz-Spohn (KLS) lattice gas model \cite{katz84, hager01a,Krapivsky13a}, which features particle diffusion under on-site exclusion and nearest-neighbors interactions and is described by a nonlinear diffusivity with a sharp maximum and a mobility with a local minimum, see Fig.~\ref{fig2}.(a). 

\begin{figure}
\vspace{-5mm}
\includegraphics[width=\linewidth]{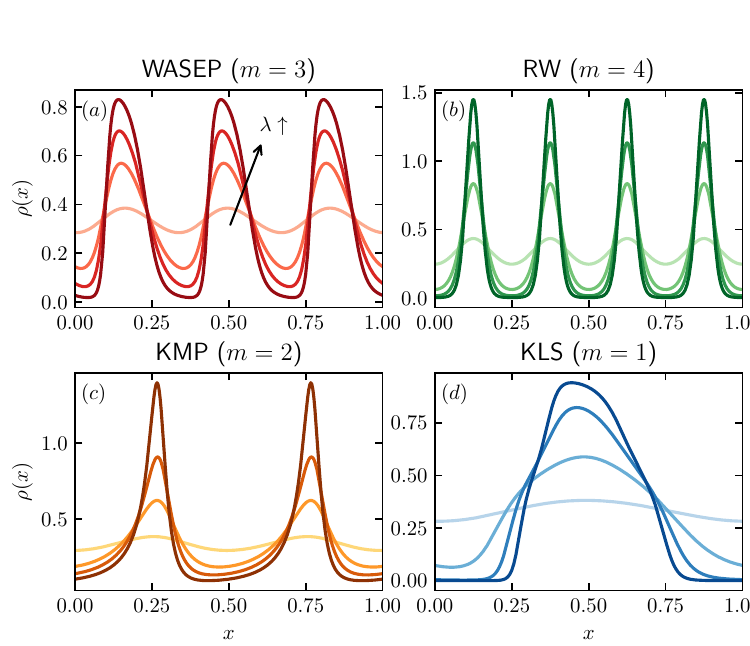}
\vspace{-5mm}
\caption{
Condensates density profiles for different models, order $m$ and couplings $\lambda$. (a) WASEP with $\Eorder=3$, (b) RW fluid with $\Eorder=4$, (c) KMP model with $\Eorder=2$, and (d) KLS lattice gas with $\Eorder=1$. In all cases $\meandens = 1/3$, $\Ed=10~\Eorder$, and $\Epampl/\Epamplcrit{\Eorder} = 1.01,~1.2,~1.5,~2$.
}
\label{fig3}
\end{figure}

All these models, introduced in more detail in Appendix \ref{app3}, exhibit programmable time-crystal phases with novel critical properties. To show this, we solved numerically Eq.~\eqref{eq:hydro_field} using the prescribed $D(\rho)$ and $\sigma(\rho)$ in each case, see Appendix \ref{app4}. Fig.~\ref{fig2} shows results for $m=1$, $\epsilon=10$ and different values of $\rho_0$ for each model. In particular, Fig.~\ref{fig2}.(b) shows the magnitude of the packing order parameter $|z_1|$ as a function the coupling $\lambda$. As anticipated above, the homogeneous density profile becomes unstable for $\lambda>\lambda_c^{(1)}$, as signaled by the packing of the density field ($|z_1|\ne 0$) around an emergent localization center, and a phase transition to a time-crystal phase in the form of a traveling condensate takes place. Interestingly, the transition for the RW, KMP and WASEP models is continuous $\forall \rho_0$, see Fig.~\ref{fig2}.(b). The velocity $\omega_1$ of the condensate in these models is initially proportional to $\sigma'(\rho_0)$ and $\epsilon$, as predicted, see Fig.~\ref{fig2}.(c), and depends monotonously on the coupling $\lambda$. This implies, in particular, a constant positive velocity for the RW model and a positive condensate velocity in the KMP model increasing with $\lambda$. For the WASEP, the sign of $\omega_1$ changes across $\rho_0=1/2$ due to its particle-hole symmetry: a particle condensate moving to the left for $\rho_0>1/2$ can be seen as a hole condensate moving to the right, and viceversa. The excess current $J-J_0$ right after the instability depends instead on the value of $\sigma''(\rho_0)$, see Fig.~\ref{fig2}.(d), so that $J>J_0$ for the KMP ($\sigma''(\rho_0)>0$ $\forall \rho_0$) and $J<J_0$ for the WASEP ($\sigma''(\rho_0)<0$ $\forall \rho_0$), while $J=J_0$ for the non-interacting RW fluid.

Results for the KLS model are more intriguing due to the change of convexity in its mobility and the sharp maximum in the diffusivity, see Fig.~\ref{fig2}.(a). In particular, for $\meandens=1/3$ the KLS lattice gas has $\sigma''(\rho_0)<0$ and it qualitatively behaves as the WASEP, at least close to the transition. Indeed the transition is continuous, the condensate velocity in positive ($\sigma'(\rho_0)>0$), and the excess current is negative ($\sigma''(\rho_0)<0$), see blue dashed lines in Figs.~\ref{fig2}.(b)-(d), though the KLS condensate velocity $\omega_1(\lambda)$ is non-monotonous and exhibits a maximum at a coupling $\lambda\approx 1.25~\lambda_c^{(1)}$, see Fig.~\ref{fig2}.(c). However, the nature of the KLS time-crystal transition changes radically for $\rho_0=1/2$, where both the order parameter and the excess current present an abrupt, discontinuous change accompanied by a region of bistability and hysteresis for $0.75 \lesssim \lambda/\lambda_c \le 1$, see Figs.~\ref{fig2}.(b),(d), all trademarks of a first-order phase transition. This bistable, first-order-type behavior stems from the pronounced peak in $\diffcoef(\dens)$ at $\dens = 1/2$, see inset to Fig.~\ref{fig2}.(a). Indeed, the stability of the condensate depends on the competition between diffusion, washing out any structure, and the packing field (proportional to the packing order parameter), reinforcing the condensate. In the homogeneous phase for $\lambda \lesssim \lambda_c$ we have $\rho(x,t)\approx 1/2~\forall x$, so a low density packing competes with almost maximal diffusivity all across the system, difficulting the condensate emergence. On the contrary, if for the same $\lambda$ we start from a condensate profile where $\rho(x,t)\ne 1/2$ almost everywhere (except at a sharp region around the condensate walls), we expect a high packing field competing with a low diffusivity, enhancing the condensate stability. This dual behavior explains the first-order scenario observed numerically, and suggests that any model with one or several sharp maxima in diffusivity may exhibit similar phenomenology. Note also that the condensate velocity is zero $\forall \lambda$ at $\rho_0=1/2$ due to the particle-hole symmetry of the KLS model \cite{perez-espigares13a}.

\begin{figure}
\centering
\includegraphics[width=1\linewidth]{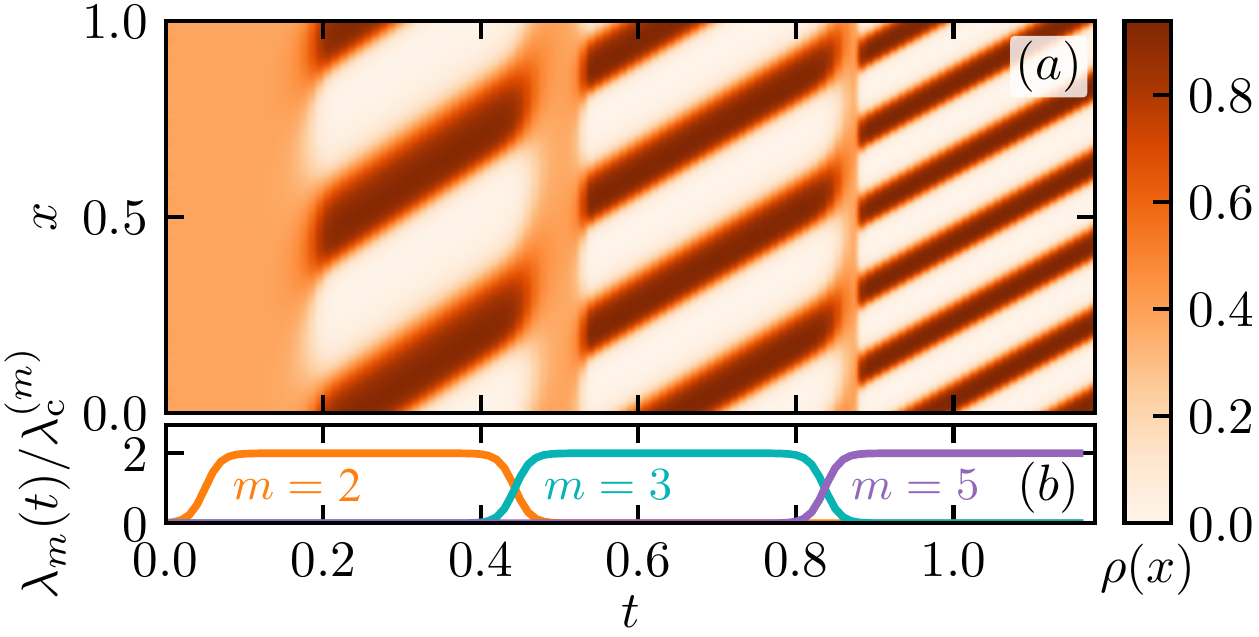}
\includegraphics[width=1\linewidth]{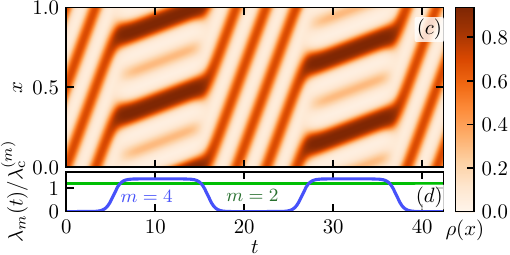}
\vspace{-5mm}
\caption{
Raster plots of the spatiotemporal evolution of the density field in the WASEP subject to different time-modulated generalized external fields $E_{x,t}[\rho]$ for $\meandens=1/3$. In (a) we swap for $\Ed=10$ between different number of condensates in time by switching on and off different orders $m=2,~3,~5$ modulating $\lambda_m(t)$ as shown in panel (b). In (c) a decorated time-crystal phase emerges for $\Ed=0.5$ by modulating in time a higher-order $m=4$ mode using $\lambda_{4}(t)$ as in panel (d), in a constant background $m=2$ matter wave obtained by setting $\lambda_2>\lambda_c^{(2)}$. 
}
\label{fig4}
\end{figure}

Fig.~\ref{fig3} shows the condensate density profiles obtained numerically for the different models for $\rho_0=1/3$, varying orders $m=1,~2,~3,~4$ and several supercritical couplings $\lambda>\lambda_c^{(m)}$. Interestingly, the shape of the condensate in each case reflects the nonlinear transport properties of the model at hand. For WASEP, the current in the time-crystal phase is lower than in the homogeneous phase due to the exclusion interaction (Fig.~\ref{fig2}.(d), $J/J_0<1$), meaning that the emergence of condensates jams dynamics on average. This jamming gives rise in turn to a sharp density accumulation at the tail of the condensate, see Fig.~\ref{fig3}.(a), while the condensate front displays a soft decay as expected due to the available free space. For the KMP heat transport model the picture is complementary: the excess current is positive ($J/J_0>1$), dynamics in the time-crystal phase is faster than in the homogeneous phase, and condensates thus exhibit a sharp front and a softer tail, see Fig.~\ref{fig3}.(c). On the other hand, the linearity of the RW fluid implies a completely symmetric condensate shape [Fig.~\ref{fig3}.(b)], while the KLS highly nonlinear transport coefficients are reflected in a intricate condensate shape, see Fig.~\ref{fig3}.(d), with WASEP-like behavior in high-$\rho$ and low-$\rho$ regions where $\sigma''(\rho)<0$, and KMP-like shape at intermediate densities where $\sigma''(\rho)>0$. We also note that the driving field $\epsilon$ controls both the velocity of the resulting condensates and the asymmetry of density profiles (not shown).

The observed time-crystal phases can be further enhanced with higher-order matter waves by introducing competing packing fields modulated in time. As a proof of concept, let us consider a generalized external field $E_{x,t}[\rho] = \epsilon + \sum_m \lambda_m(t) \mathcal{E}_x^{(m)}[\rho]$. Fig.~\ref{fig4} displays the spatiotemporal evolution of the density field that results from the numerical integration of Eq.~\eqref{eq:hydro_field} subject to different generalized external fields (see Appendix \ref{app4}). For instance, we may swap between different number of condensates in time as shown in Fig.~\ref{fig4}.(a) by switching on and off different orders $m$ modulating $\lambda_m(t)$ as in Fig.~\ref{fig4}.(b). We may also obtain custom decorated time-crystal phases by switching on and off in time a higher-order $2m$ mode using $\lambda^{(2m)}(t)$ as in Fig.~\ref{fig4}.(d), in a constant background matter wave obtained by setting $\lambda^{(m)}>\lambda_c^{(m)}$. Interestingly, a time-dependent decorated pattern emerges, switching in-phase with $\lambda_{2m}(t)$ between a symmetric time-crystal phase with $m$ condensates when $\lambda_{2m}(t)\approx 0$ and $2m$ asymmetric condensates when $\lambda_{2m}(t)> \lambda_c^{(2m)}$. These examples, just two among a myriad of interesting combinations, showcase the potential of the packing-field route to engineer and control programmable time-crystal phases in driven diffusive fluids, opening new avenues of future research with promising technological applications.

\section{Discussion and outlook}
\label{sec:5}
Interestingly, the packing field~\eqref{eq:extfield_pairs_field} can be written as a generalized Kuramoto-like long-range interaction \cite{hurtado-gutierrez20a,kuramoto87a,pikovsky03a}, in which case Eq.~\eqref{eq:hydro_field} resembles the one for the oscillator density in the mean-field Kuramoto synchronization model \cite{acebron05a}. It is hence tempting to relate the \emph{explosive} time-crystal phase transition observed here for the KLS lattice gas with similar first-order synchronization transitions reported in certain oscillator models \cite{leyva12a,Hu14a}. These links are only formal however, as synchronization models lack any transport in real space, while in our case the nonlinearity of diffusion and mobility coefficients caused by local exclusion and interactions introduces crucial differences in the observed custom time-crystal phases.

In this work we have shown how to leverage the concept of packing field~\eqref{eq:extfield_pairs_field} to engineer programmable continuous time crystals in driven diffusive fluids, characterized by multiple rotating condensates. These phases can be created in the lab with current technology, using for instance assemblies of colloidal particles confined in ring-shaped light traps, created e.g. with infrared optical tweezers rapidly steered through a acousto-optic deflector \cite{cereceda-lopez23a,cereceda-lopez23b,cereceda-lopez24a}; see also \cite{lutz04a,villada-balbuena21a}. Packing fields could be implemented via a feedback loop from real-time particle tracking, by modulating the depth of the individual lattice traps to bias motion locally. The challenge remains to exploit this route to time crystals in this and other geometries.

\begin{acknowledgments}
The research leading to these results has received funding from the I+D+i grants PID2023-149365NB-I00, PID2020-113681GB-I00, PID2021-128970OA-I00, C-EXP-251-UGR23 and  P20\_00173, funded by MICIU/AEI/10.13039/501100011033/, ERDF/EU, and Junta de Andaluc\'{\i}a - Consejer\'{\i}a de Econom\'{\i}a y Cono-cimiento, as well as from fellowship FPU17/02191 financed by the Spanish Ministerio de Universidades. We are also grateful for the the computing resources and technical support provided by PROTEUS, the supercomputing center of Institute Carlos I in Granada, Spain.
\end{acknowledgments}

\appendix

\section{Hydrodynamic instability in the time-crystal phase transition}
\label{app1}

Our starting point is the hydrodynamic evolution equation for the density field $\rho(x,t)$ in a $1d$ periodic diffusive system driven by an external field $E_x[\rho]$,
\begin{equation}
\partial_t \rho = - \partial_x \Big[-D(\rho) \partial_x \rho + \sigma(\rho) E_x[\rho] \Big] \, ,
\label{eq:hydro_fieldAPP}
\end{equation}
with $\pos \in [0, 1]$, and $\diffcoef(\dens)$ and $\mobility(\dens)$ the diffusivity and the mobility transport coefficients, respectively. The external field takes the form $E_x[\rho] = \epsilon +  \lambda \mathcal{E}_x^{(m)}[\rho]$, where $\epsilon$ is a constant driving and $\lambda$ is the coupling to a $m$-th order packing field $\mathcal{E}_x^{(m)}[\rho]$, defined as 
\begin{equation}
\mathcal{E}^{(m)}_x[\rho] = \frac{1}{\rho_0} \int_0^1 \dd y \, \rho(y,t) \, \sin\left(2\pi m (y-x) \right) \, ,
\label{eq:extfield_pairs_fieldAPP}
\end{equation}
where $\rho_0=\int_0^1 \rho(x, t) \dd x$ is the conserved average density. To better understand the action of $\mathcal{E}_x^{(m)}[\rho]$, we define now the complex $\Eorder$th-order packing order parameter (also known as the Kuramoto-Daido order parameter in the context of synchronization transitions),
\begin{equation}
z_m[\rho] = \frac{1}{\rho_0} \int_0^1 dx \, \rho(x,t) \, \textrm{e}^{\textrm{i}2\pi m x} \equiv \abs{z_m} \textrm{e}^{\textrm{i} \varphi_m}\, .    
\label{eq:z_fieldAPP}
\end{equation}
Its magnitude $\abs{\oparm}$ measures the packing of the density field around $\Eorder$ equidistant \emph{emergent localization centers} placed at angular positions $\phi_m^{(j)}=\arg[(\hspace{-3pt}\sqrt[m]{\oparm})_j]=(\varphi_m + 2\pi j)/m$, with $j\in[0,m-1]$. Using $z_m[\rho]$, the packing field of Eq.~\eqref{eq:extfield_pairs_fieldAPP} can be simply rewritten as $\mathcal{E}_x^{(m)}[\rho] =  \abs{z_m} \sin(\varphi_m - 2\pi\pos \Eorder)$. In this way, $\mathcal{E}^{(m)}_x[\rho]$ drives particles locally towards the $\Eorder$ emergent localization centers placed at $\phi_m^{(j)}\in[0,2\pi)$,  pushing particles that lag behind the closest localization center while restraining those moving ahead, with an amplitude proportional to the amount of local packing as measured by $\abs{\oparm}$. This results in a nonlinear feedback mechanism that amplifies the local packing fluctuations naturally present in the system, resulting eventually in the emergence of $\Eorder$ traveling-wave condensates for large enough values of $\Epampl$, and exhibiting the fingerprints of spontaneous time-translation symmetry breaking.

To determine the critical threshold $\Epampl_c^{(\Eorder)}$ for this instability to happen, we first note that for any value of $\Epampl$ the homogeneous density profile $\densxt = \meandens$ is a solution of the hydrodynamic equation \eqref{eq:hydro_fieldAPP}. Therefore, a linear stability analysis of this solution will allow us to find the critical value $\Epampl_c^{(\Eorder)}$. We hence consider a small perturbation over the flat profile, $\densxt = \meandens + \denspxt$, with $\int_0^{1} d\pos \denspxt = 0$ so as to conserve the global density of the system. Plugging this perturbation into Eq.~\eqref{eq:hydro_fieldAPP} and linearizing it to first order in $\denspxt$ we obtain
\begin{eqnarray}
\partial_t \delta\rho = &-&\partial_x \Big[-D(\rho_0) \partial_x\delta\rho + \epsilon \sigma'(\rho_0) \delta\rho + \\ 
&+&\sigma(\rho_0) \big(\epsilon+\lambda \abs{\oparm[\delta\dens]} \sin(\varphi_m[\delta\dens] - 2\pi\pos \Eorder) \big) \Big] \, ,\nonumber
\label{eq:hydro_linearized_genpackfield}
\end{eqnarray}
where $\mobility'(\rho_0)$ stands for the derivative of the mobility $\mobility(\dens)$ with respect to its argument evaluated at $\rho_0$, and where we have used that $\abs{\oparm}$ is already first-order in $\delta\rho$, see Eq. \eqref{eq:z_fieldAPP}. The system periodicity can be used to expand the density field perturbation in Fourier modes,
\begin{equation}
    \denspxt = \sum_{j=-\infty}^{\infty} \fcoeft{j} e^{i 2\pi\pos j} \, ,
\end{equation}
where the $j$-th Fourier coefficient is given by $\fcoeft{j} = \int_0^{1} \dd x\denspxt \textrm{e}^{-\textrm{i} 2\pi\pos j}$.
Noting that the Kuramoto-Daido parameter is proportional to the $(-m)$-th Fourier coefficient in this expansion, i.e. $\oparm[\delta\dens] = \fcoeft{-\Eorder} / \meandens$, and replacing the Fourier expansion in Eq.~\eqref{eq:hydro_linearized_genpackfield}, we obtain
\begin{equation}
\sum_{j=-\infty}^{\infty}\Big(\partial_t C_j(t) + \zeta_j C_j(t) \Big) \textrm{e}^{\textrm{i} 2\pi\pos j } = 0 \, ,
\label{Fourier1}
\end{equation}
where we have defined
\begin{equation}
\label{Fourier2}
\zeta_j \equiv \left(2\pi j\right)^2 \diffcoef (\meandens)  + \textrm{i} 2\pi j \mobility^{\prime}(\meandens)\Ed  - \Epampl\frac{\mobility(\meandens)}{2\meandens} 2\pi\Eorder (\delta_{j, \Eorder} + \delta_{j,-\Eorder}) \, ,
\end{equation}
and $\delta_{j, \Eorder}$ and $\delta_{j,-\Eorder}$ are Kronecker deltas. As the different complex exponentials in Eq. \eqref{Fourier1} are linearly independent, each parenthesis in the equation must be zero. Therefore the solution for the different Fourier coefficients is just 
\begin{equation}
\fcoeft{j} = C_j(0) \textrm{e}^{-\zeta_j \tv} \, ,
\end{equation}
with $C_j(0)$ the coefficients associated with the initial perturbation $\delta\dens(x,0)$. The stability of the different Fourier modes is then controlled by the real part of $\zeta_j$, for which we have to consider two distinct cases: $\abs{j} \neq \Eorder$ and $\abs{j} = \Eorder$. In the first case $\abs{j} \neq \Eorder$, we have $\Re(\zeta_j) = \diffcoef(\meandens) \left(2\pi j\right)^2>0\; \forall j$, so that these Fourier modes will always decay. On the other hand, when $\abs{j} = \Eorder$, the decay rate involves a competition between the diffusion term and the packing field, 
\begin{equation}
\label{Recrit}
\Re(\zeta_{\pm m}) = \left(2\pi \Eorder\right)^2\left( \diffcoef(\meandens)  - \Epampl\frac{\mobility(\meandens)}{4\pi \Eorder\meandens}\right) \,.
\end{equation}
The critical value of $\Epampl$ is reached whenever $\Re(\zeta_{\pm m}) = 0$, and reads
\begin{equation}
\label{eq:clambda_genpackfield}
\Epamplcrit{\Eorder} = 4\pi\Eorder \frac{\diffcoef(\meandens)\meandens}{\mobility(\meandens)} \, .
\end{equation}
In this way we expect the homogeneous density solution $\densxt = \meandens$ to become unstable for $\lambda>\lambda_c^{(\Eorder)}$, leading to a density field solution with a more intricate spatiotemporal structure. 

The previous analysis shows that, right after the instability, the first modes to become unstable and contribute to a structured density field will be the $\pm\Eorder$-order Fourier modes. In this regime we therefore expect a traveling-wave solution $\dens(x,t)=\meandens + A\cos(\omega_m t - 2\pi\Eorder\pos)$ with $A$ a small amplitude and where the angular velocity $\omega_m$ is given from the imaginary part of Eq.~\eqref{Fourier2},
\begin{equation}
\label{eq:omega_twperturbative_genpackfield}
\omega_m = 2\pi \Eorder \mobility'(\meandens)\Ed \, .
\end{equation}
This suggests that beyond the instability, the homogeneous density turns into $m$ condensates periodically moving at a constant velocity $\omega_m$ (initially proportional to $\mobility'(\meandens)$ and $\Ed$), thus giving rise to a custom continuous time crystal. Moreover, the average current $\meancurrent=\int_0^1 \dd\pos j(x,t)$ associated to this traveling-wave solution right after the instability can be calculated from the local current in the linearized equation \eqref{eq:hydro_linearized_genpackfield}, resulting in
\begin{equation}
\label{eq:J_twperturbative_genpackfield}
\meancurrent = \meancurrentflat+ A^2 \mobility''(\meandens) \Ed/4 \, ,
\end{equation}
where $\meancurrentflat = \mobility(\meandens) \Ed$ is the average current in the homogeneous phase. While we only expect Eqs.~\eqref{eq:omega_twperturbative_genpackfield}-\eqref{eq:J_twperturbative_genpackfield} to hold true close to $\Epampl = \Epamplcrit{\Eorder}$, they highlight the relevance of the transport coefficients in the response of the model to the packing field. Depending on the slope and convexity of the mobility of each model, the packing field will enhance or lower the current and the speed of the resulting condensates.

\section{Mapping condensate profiles across different packing orders}
\sectionmark{Mapping condensate profiles across different packing orders}
\label{app2}
    
In this section we investigate the relation between traveling-wave solutions to Eq.~\eqref{eq:hydro_fieldAPP} corresponding to different packing orders $\Eorder$. In particular we will show that, provided that a $2\pi/m$-periodic traveling-wave solution of Eq.~\eqref{eq:hydro_fieldAPP} exists for packing order $\Eorder$, this solution can be built by gluing together $\Eorder$ copies of the solution of the $m=1$ equation with properly rescaled parameters. To prove this, we start by using a traveling-wave ansatz $\dens(\pos,\tv) = \denstw{\Eorder}(\omega \tv - 2\pi \pos)$ in the hydrodynamic equation \eqref{eq:hydro_fieldAPP} with packing order $\Eorder$, where $\denstw{\Eorder}$ is a generic periodic function and $\omega$ denotes the traveling-wave velocity. This leads to
\begin{eqnarray}
\label{eq:hydrotw_genpackfield}
\omega \denstw{\Eorder}^{\prime}(\twarg) &=&2\pi \dv{\twarg} \Big\{ D(\denstw{\Eorder}) 2\pi \denstw{\Eorder}^{\prime}(\twarg) + \\
&&+\mobility(\denstw{\Eorder}) \big[\Ed + \Epampl \abs{\oparm} \sin(\oparargmini + \Eorder\twarg)\big]\Big\}\, ,\nonumber
\end{eqnarray}
where we have introduced the variable $\twarg = \omega \tv - 2\pi \pos$. In addition, we have used that under the traveling-wave ansatz the magnitude of the order parameter is constant and its complex phase increases linearly in time, i.e., $\oparm[\denstw{\Eorder}] = \abs{\oparm} \exp(i\oparargm[\denstw{\Eorder}])$ with $\oparargm[\denstw{\Eorder}]=\oparargmini + \Eorder\omega t$. 

Since we expect the formation of $m$ equivalent particle condensates once the homogeneous density profile becomes unstable, it seems reasonable to assume that the resulting traveling density wave will exhibit $(2\pi/\Eorder)$-periodic behavior, i.e. $\denstw{\Eorder}(\twarg) = \denstwt(\Eorder\twarg)$ with $\denstwt$ a new $2\pi$-periodic function. Under this additional assumption, the initial $\Eorder$-th order Kuramoto-Daido parameter reads
\begin{eqnarray}
\oparm(0) &=& \frac{1}{\meandens} \int_0^{1} \dd \pos \denstw{\Eorder}(- 2\pi \pos) \mathrm{e}^{i 2\pi\Eorder\pos} =\\ \nonumber
&=&\frac{1}{\meandens} \int_0^{1} \dd \pos \denstwt(- 2\pi \Eorder \pos) \mathrm{e}^{i 2\pi\Eorder\pos} = \\ \nonumber
&=&\frac{1}{\meandens} \int_0^{1} \dd \tilde{\pos} \denstwt(- 2\pi \tilde{\pos}) \mathrm{e}^{i 2\pi\tilde{\pos}} \equiv {\opar}_1(0) \, , 
\end{eqnarray}
where we haven taken into account the periodicity of $\denstwt$ and where ${\opar}_1(0)$ is defined as the initial $m=1$ Kuramoto-Daido parameter of $\denstwt(\twarg)$. Using this result we can rewrite Eq.~\eqref{eq:hydrotw_genpackfield} in terms of $\denstwt$ and a new variable $\tilde{\twarg} = \Eorder \twarg$,
\begin{eqnarray}
\frac{\omega}{\Eorder} \denstwtprime(\twargt) &=& 2\pi \dv{\twargt} \Big\{D(\denstwt) 2\pi \denstwtprime(\twargt) + \\ 
&&+ \mobility(\denstwt) \big[\frac{\Ed}{\Eorder} + \frac{\Epampl}{\Eorder} \abs{\tilde{\opar}_1} \sin(\tilde{\varphi}_{1,0} - \twargt)\big]\Big\} \, , \nonumber
\label{wave2}
\end{eqnarray}
where we have used that $\denstw{\Eorder}^{\prime}(\twarg) = \Eorder \denstwtprime(\twargt)$. This is nothing but the original equation for the traveling wave Eq.~\eqref{eq:hydrotw_genpackfield} with a packing order $\Eorder=1$ and rescaled parameters
\begin{equation}
\label{eq:params_rescaling_profile_equivalence}
\tilde{\omega} = \frac{\omega}{\Eorder}  \, , \qquad  \tilde{\Ed} = \frac{\Ed}{\Eorder} \, , \qquad \tilde{\Epampl}=\frac{\Epampl}{\Eorder} \, .
\end{equation}
In this way, we have proved that if $\denstwt(\tilde{\omega} \tv - 2\pi\pos)$ is a traveling-wave solution of the hydrodynamic equation \eqref{eq:hydro_fieldAPP} with $m=1$ and parameters $\tilde{\Ed}$ and $\tilde{\Epampl}$, then $\dens(\pos, \tv) = \denstwt(\Eorder\tilde{\omega} \tv - \Eorder2\pi\pos)$ is a solution of the corresponding hydrodynamic equation with order $\Eorder$ and parameters $\Ed=\Eorder\tilde{\Ed}$ and $\Epampl = \Eorder \tilde{\Epampl}$. Note that this result resembles the one found in the Kuramoto model, where a complete dynamical equivalence between the first-order and higher-order couplings has been reported \cite{delabays2019}.

\section{Driven diffusive models}
\label{app3}

The general results obtained in this paper have been illustrated by solving the hydrodynamic equations for several paradigmatic driven diffusive systems, including the random walk (RW) fluid \cite{spohn12a}, the weakly asymmetric simple exclusion process (WASEP) for interacting particle diffusion \cite{spitzer70a,derrida98a}, the Kipnis-Marchioro-Presutti (KMP) heat transport model \cite{kipnis82a}, and the Katz-Lebowitz-Spohn (KLS) lattice gas \cite{katz84,hager01a,Krapivsky13a}. In this appendix we briefly introduce these microscopic lattice models and their hydrodynamic description. We define the microscopic models on a $1d$ lattice of size $L$ with periodic boundary conditions, though they can be easily generalized to arbitrary dimension and different boundary conditions. As for their hydrodynamic description, it takes in all cases a standard diffusive form for a mesoscopic density field $\rho(x,t)$,
\begin{equation}
\partial_t \rho = - \partial_x \Big[-D(\rho) \partial_x \rho + \sigma(\rho) E \Big] \, ,
\label{eq:hydro_fieldAPP0}
\end{equation}
with $\pos \in [0, 1]$, $\diffcoef(\dens)$ and $\mobility(\dens)$ the diffusivity and the mobility transport coefficients, respectively, and $E$ some external  field that drives the system to a nonequilibrium steady state of global density $\rho_0=\int_0^1 \rho(x,t)~dx$ and a net current $\langle q\rangle=\sigma(\rho_0) E$.

The RW fluid is composed by $N$ \emph{independent} particles which jump stochastically and sequentially to nearest neighbor lattice sites with rates $r_\pm=\frac{1}{2}\exp(\pm E/L)$ for jumps along the $\pm \hat{x}$-direction, with $L$ the lattice size such that $\rho_0=N/L$. At the hydrodynamic level, the RW fluid is characterized by a diffusivity $D(\rho)=1/2$ and a mobility $\sigma(\rho)=\rho$ \cite{spohn12a}. This linear dependence of the mobility on the density field is a signature of the noninteracting character of the RW fluid.

The weakly asymmetric simple exclusion process (WASEP) \cite{spitzer70a,derrida98a} is a stochastic particle system similar to the RW fluid, but with the crucial addition of a exclusion interactions. In particular, in the WASEP $N$ particles live in a periodic $1d$ lattice of size $L$, such that each lattice site may contain at most one particle. Dynamics is stochastic and proceeds via sequential particle jumps to nearest-neighbor sites, provided these are empty (in other case the exclusion interaction forbides the jump), at a rate $r_\pm=\frac{1}{2}\exp(\pm E/L)$ for jumps along the $\pm \hat{x}$-direction. At the macroscopic level the WASEP is characterized by a diffusivity $D(\rho)=1/2$ and a mobility $\sigma(\rho)=\rho (1-\rho)$, a quadratic dependence on the local density clearly reflecting the key role of the exclusion interaction.

In the Kipnis-Marchioro-Presutti (KMP) model of heat transport \cite{kipnis82a,hurtado09c,hurtado11a}, each lattice site $i\in[1,L]$ is characterized by a non-negative amount of energy $\rho_i\ge 0$. Dynamics is stochastic, proceeding through random energy exchanges between randomly chosen nearest neighbors $(i,i+1)$, in such a way that the total pair energy is conserved in the collision. At the hydrodynamic level, the KMP model is characterized by a diffusivity $D(\rho)=1/2$ and a mobility $\sigma(\rho)=\rho^2$.

The Katz-Lebowitz-Spohn (KLS) lattice gas model is a stochastic particle systems that features on-site exclusion and nearest-neighbor interactions \cite{katz84,hager01a,Krapivsky13a,baek15a}. In the KLS model each lattice site can be empty or occupied by one particle at most. The model is defined by two parameters, $\delta$ and $\eta$, which control the particle hopping dynamics with rates $0100 \xrightarrow{1+\delta} 0010$, $1101 \xrightarrow{1-\delta} 1011$, $1100 \xrightarrow{1+\eta} 1010$, and $0101 \xrightarrow{1-\eta} 0011$. Note that spatially inverted versions of these transitions occur with identical rates \cite{baek15a}.  In contrast to the other microscopic transport models presented above, the richer dynamics of the KLS model leads to more complex transport coefficients at the macroscopic level. Specifically, the diffusion coefficient is obtained in terms of the quotient
\begin{align}
D(\rho) = \frac{\mathcal{J}(\rho)}{\chi(\rho)} \,,
\end{align}
where $\mathcal{J}(\rho)$ is the average current in the totally asymmetric version of the model and $\chi(\rho)$ is its compressibility.
The first is given by
\begin{equation}
\mathcal{J}(\rho) = \frac{\nu [1+\delta(1-2\rho)]-\eta\sqrt{4\rho(1-\rho)}}{\nu^3} \,,
\end{equation}
while the second obeys,
\begin{equation}
\chi(\rho) = \rho(1-\rho)\sqrt{(2\rho-1)^2 + 4\rho(1-\rho)e^{-4\beta}}\, .
\end{equation}
Parameters $\nu$ and $\beta$ are determined in turn from the expressions
\begin{equation}
\nu \equiv \frac{1+ \sqrt{(2\rho-1)^2 + 4\rho(1-\rho)e^{-4\beta}}}{\sqrt{4\rho(1-\rho)}} \, , \qquad e^{4\beta} \equiv \frac{1+\eta}{1-\eta}\, ,
\end{equation}
Finally the mobility coefficient $\sigma(\rho)$ is obtained from the diffusion coefficient and the compressibility using the Einstein relation $\sigma(\rho) = 2 D(\rho) \, \chi(\rho)$. For this paper we have chosen to work with parameters $\eta=0.9$ and $\delta=0$, which results in a nonlinear diffusivity with a sharp maximum and a mobility with a local minimum, see Fig.~\ref{fig2}.(a) in the main text.

\section{Solving numerically the traveling-wave hydrodynamical equation}
\label{app4}

\newcommand{\inicond}{y_0}
\newcommand{\odeconst}{C}

In this appendix, we detail the numerical method used in this work to calculate the traveling-wave solutions to Eq.~\eqref{eq:hydro_fieldAPP}, i.e.
\begin{eqnarray}
\label{eq:hydroeq_ALL}
\partial_t \dens = - \partial_x \Big[&-&\diffcoef(\rho) \partial_x \dens + \mobility(\dens) \Big(\Ed +  \\ 
&+&\Epampl \abs{\oparm[\dens]} \sin(\oparargm[\dens] - 2\pi\pos \Eorder)\Big) \Big] \, , \nonumber
\end{eqnarray}
with $\pos \in [0, 1]$, periodic boundary conditions, and $\oparm[\dens]$ the Kuramoto-Daido parameter given by 
\begin{align}
\label{eq:opar_selfconsisthydro}
\oparm[\dens] &= \frac{1}{\meandens} \int_0^1 \dd x \rho(x,t) \expsup{\textrm{i}2\pi \Eorder x} \equiv \abs{\oparm[\dens]} \expsup{\textrm{i} \oparargm[\dens]} \, .
\end{align}
This is a nonlinear second-order integro-differential equation that poses a challenge for standard numerical methods. In particular, reaching the traveling-wave regime with enough precision using standard techniques for partial differential equations---such as finite difference methods--- becomes increasingly difficult when the system is either close to the critical point or deep into the nonlinear regime. 

To address this problem, we have devised an alternative approach based on transforming this equation into an ordinary first-order differential equation supplemented by several self-consistence relations. For that, we consider the hydrodynamic equation \eqref{eq:hydroeq_ALL} with a traveling-wave ansatz $\dens(x,t) = \denstw{\Eorder}(\omega t - 2\pi x)$ to obtain an ordinary second-order differential equation in terms of the variable $\twarg = \omega t - 2\pi x$,
\begin{eqnarray}
\label{eq:hydrotw_selfconsisthydro}
\omega \denstw{\Eorder}^{\prime}(\twarg) &=&2\pi \dv{\twarg} \Big\{ \diffcoef(\denstw{\Eorder}) 2\pi \denstw{\Eorder}^{\prime}(\twarg) + \\
&&+\mobility(\denstw{\Eorder}) \big[\Ed + \Epampl \abs{\oparm} \sin(\oparargmini + \Eorder\twarg)\big]\Big\}  \, . \nonumber
\end{eqnarray}
Here we have used that, under the traveling wave ansatz, the magnitude of the order parameter is constant and its complex phase increases linearly in time, i.e., $\oparm[\denstw{\Eorder}] = \abs{\oparm} \expsup{i(\oparargmini + \Eorder\omega t)}$ with $\oparargmini$ the argument at $t=0$. This simplifies the equation on one hand, but it makes it harder to deal with the order parameter. In the original partial differential equation, given the density profile at a particular time step, we just needed to evaluate $\oparm(\dens)$ in order to obtain the next one. However, in the traveling-wave version of Eq.~\eqref{eq:hydrotw_selfconsisthydro}, in order to compute the differential equation we need the integral Eq.~\eqref{eq:opar_selfconsisthydro} of its solution, which renders usual differential equations methods invalid.

To overcome this issue, we first integrate Eq.~\eqref{eq:hydrotw_selfconsisthydro} to obtain a first-order differential equation easier to tackle numerically,
\begin{eqnarray}
\label{eq:hydrotwfirst_selfconsisthydro}
\omega \denstw{\Eorder}(\twarg) &=& 2\pi \Big\{ D(\denstw{\Eorder}) 2\pi \denstw{\Eorder}^{\prime}(\twarg) + \\
&&+\mobility(\denstw{\Eorder}) \big[\Ed + \Epampl \abs{\oparm} \sin(\Eorder\twarg)\big]\Big\} + \odeconst \, , \nonumber
\end{eqnarray}
where $C$ is an integration constant and we have chosen $\oparargmini = 0$ without loss of generality (i.e. we set the origin of $\twarg$ to the angular position given by $\oparargmini$, so $\twarg \mapsto \twarg - \oparargmini$). The key step now is to consider $|\oparm|$ as a free parameter instead of the integral of the solution, thus transforming the previous equation into a standard ordinary differential equation depending on three parameters: $|\oparm|$, $\omega$, and $\odeconst$. Such a differential equation can now be solved using the initial condition at the left boundary $\denstw{\Eorder}(0) = \inicond$ to obtain the solution $\denstw{\Eorder}(\twarg; \inicond, |\oparm|, \omega, \odeconst)$. Nevertheless, this function is not in general a solution to the original problem. The parameters $\inicond$, $|\oparm|$, $\omega$, and $\odeconst$ must be carefully chosen to ensure the compatibility of the solution with the specifications of the original problem: the periodicity of the solution, its average density $\meandens$ and the consistency between the chosen $|\oparm|$ and the value calculated from $\denstw{\Eorder}(\twarg; \inicond, |\oparm|, \omega, \odeconst)$. These conditions are captured in the following system of equations,
\begin{align}
    \denstw{\Eorder}(0; \inicond, |\oparm|, \omega, \odeconst)
    =
    \denstw{\Eorder}(2\pi; \inicond, |\oparm|, \omega, \odeconst)
    \,, 
    \\
    \meandens
    =
    \int_0^1 \dd x \denstw{\Eorder}(-2\pi x; \inicond, |\oparm|, \omega, \odeconst)
    \,, 
    \\
    \abs{\oparm}
    =
    \int_0^{1} \dd x \denstw{\Eorder}(-2\pi x; \inicond, |\oparm|, \omega, \odeconst) \expsup{\textrm{i} 2\pi x \Eorder}
    \,, 
\end{align}
which complete the set of equations required to solve the problem (we have two real equations and a complex one to determine four real parameters). To determine the solution to the original problem, we define a function $G(\inicond, |\oparm|, \omega, \odeconst)$ which calculates the profile $\denstw{\Eorder}(0; \inicond, |\oparm|, \omega, \odeconst)$ using Eq.~\eqref{eq:hydrotwfirst_selfconsisthydro} and returns the squared sum of the errors in the previous self-consistent equations for this profile.
With this, the correct parameters $\inicond, |\oparm|, \omega, \odeconst$ can be found by performing a numerical optimization of $G(\inicond, |\oparm|, \omega, \odeconst)$, and the traveling wave profile $\denstw{\Eorder}$ corresponding to such parameters will be the solution of the original problem. This approach is reminiscent of the shooting method \cite{press2007} used to solve two-point boundary value problems.

\bibliography{refs}{}

\end{document}